\documentstyle[prd,aps,epsfig,amssymb]{revtex}
\let\jnfont=\rm
\def\NPB#1,{{\jnfont Nucl.\ Phys.\ B }{\bf #1},}
\def\PLB#1,{{\jnfont Phys.\ Lett.\ B }{\bf #1},}
\def\EPJC#1,{{\jnfont Eur.\ Phys.\ J.\ C }{\bf #1},}
\def\PRD#1,{{\jnfont Phys.\ Rev.\ D }{\bf #1},}
\def\PRL#1,{{\jnfont Phys.\ Rev.\ Lett.\ }{\bf #1},}
\def\MPLA#1,{{\jnfont Mod.\ Phys.\ Lett.\ A }{\bf #1},}
\def\JPG#1,{{\jnfont J.\ Phys.\ G}{\bf #1},}
\def\CTP#1,{{\jnfont Commun.\ Theor.\ Phys.\ }{\bf #1},}
\def\p_slash{\not{\hbox{\kern-2.1pt $p$}}}
\begin{document}
\draft
\preprint{}
\title{Supersymmetric effects in top quark decay into polarized W-boson}

\author{Junjie Cao $^{a,b}$,  Robert J. Oakes $^c$, Fei Wang $^b$  and Jin Min Yang $^b$ \\\ }

\address{$^a$ Department of Physics, Henan Normal University, Xinxiang, Henan 453002, China}
\address{$^b$ Institute of Theoretical Physics, Academia Sinica, Beijing 100080, China}
\address{$^c$ Department of Physics and Astronomy, Northwestern University,  Evanston, IL 60208, USA}
\date{\today}
\maketitle

\begin{abstract}

We investigate the one-loop supersymmetric QCD (SUSY-QCD) and
electroweak (SUSY-EW) corrections to the top quark decay into a
$b$-quark and a longitudinal or transverse $W$-boson. The
corrections are presented in terms of the longitudinal ratio
$\Gamma(t\to W_L b)/\Gamma(t\to W b)$ and the transverse ratio
$\Gamma(t\to W_- b)/\Gamma(t\to W b)$. In most  of the
parameter space, both SUSY-QCD and SUSY-EW corrections to these
ratios are found to be less than $1\%$ in magnitude and they tend
to have opposite signs. The corrections to the total width
$\Gamma(t\to W b)$ are also presented for comparison with the
existing results in the literature. We find that our SUSY-EW
corrections to the total width differ significantly from
previous studies: the previous studies give a large correction of
more than $10\%$ in magnitude for a large part of the parameter
space while our results reach only few percent at most.
\end{abstract}

\pacs{14.65.Ha 12.60.Jv 11.30.Pb}

\section{\bf Introduction}
As the most massive fermion in the Standard Model (SM),  the top quark
is believed to be sensitive to new physics\cite{sensitive}. Since the
measurements of top quark properties in Run~1 at the Fermilab
Tevatron have only small statistics, there remains plenty of room
for new physics to be discovered in future precision measurements.
Such possibilities exist on our horizon: while the upgraded
Tevatron collider will significantly improve the precision of the
top quark measurements, the CERN Large Hadron Collider (LHC) and
the planned Next-generation Linear Collider (NLC) will serve as
top quark factories and allow precision measurements of the top
quark properties~\cite{review}.

Although various exotic production processes \cite{new-production}
and rare decay modes \cite{new-decay} of the top quark will serve
as robust evidence for new physics (since these exotic processes
are forbidden or severely suppressed in the SM),  the role of the
dominant decay mode $t \to W b$ in probing new physics should not
be underestimated \cite{Nelson}. One advantage of this decay mode
is that it is free of non-perturbative theoretical uncertainties
\cite{Bigi} and future precision experimental data can  therefore be
compared with accurate theoretical predictions. The other advantage of
this channel is that
  the $W$-boson as decay product
is strongly
polarized and the helicity contents (transverse-plus $W_+$,
transverse-minus $W_-$ and longitudinal $W_L$) of the $W$-boson
can be probed through a measurement of the shape of the lepton
spectrum in the $W$-boson decay \cite{CDF}. Therefore, the study
of top decay into a polarized $W$-boson will provide additional
information about the $tWb$ coupling \footnote{Among the three
polarizations of the $W$-boson in top decay, the longitudinal mode
is of particular interest since it is relevant for the
understanding of the electroweak symmetry breaking mechanism.} and
has so far attracted a lot of attention from both theorists and
experimentalists. On the experimental side, the CDF collaboration
has already performed the measurement of the helicity component of
the $W$-boson in top quark decay from Run~1 data and obtained the
results
\begin{eqnarray}
&&\Gamma_L/\Gamma =0.91 \pm 0.37 (stat.) \pm 0.13 (syst.), \nonumber \\
&&\Gamma_+/\Gamma =0.11 \pm 0.15 \ ,    \nonumber
\end{eqnarray}
where $\Gamma $ is the total decay rate of $t\to Wb$, and
$\Gamma_L $ and $\Gamma_+ $ denote respectively the rates into
a longitudinal and transverse-plus $W$-boson.
Although  the error of these measurements is quite large at the
present time, it is expected to be reduced significantly during
Run~2 of the Tevatron and may reach $1\% \sim 2\%$ at the
LHC \cite{Willenbrock}. On the theoretical side,
predictions of these quantities in the SM up to one-loop level are
now available \cite{Groot1,Groot2}.
For the mass values of Refs.\cite{Groot1,Groot2},
the tree-level results are 0.703 for $\Gamma_L/\Gamma$,
0.297 for $\Gamma_-/\Gamma$ and ${\cal{O}}(10^{-4})$ for $\Gamma_+/\Gamma$,
and the QCD corrections to these predictions are respectively
$-1.07 \% $, $2.19\%$ and $0.10\%$, while the electroweak
corrections are at the level of a few per mill.

In order to probe new physics from the future precise measurement of
$\Gamma_L/\Gamma$, $\Gamma_-/\Gamma$ or  $\Gamma_+/\Gamma$,
we must know the new physics contributions to these quantities in
various models. In this article we focus on the supersymmetric contributions
in the framework of the Minimal Supersymmetric Model (MSSM).
In this model the one-loop corrections to the total width of $t \to b W$
have been studied by several groups \cite{Yang1,Yang2,Garcia,Dabelstein,Brandenburg}.
However, the corrections to $\Gamma_L/\Gamma$, $\Gamma_-/\Gamma$ or $\Gamma_+/\Gamma$
are still lacking and calculating these corrections is the major
goal of this article.

In the MSSM the genuine supersymmetric corrections
(i.e., the corrections from various sparticle loops) are
SUSY electroweak (SUSY-EW) corrections, arising from interactions
of charginos or neutralinos, and SUSY-QCD corrections,
arising from interactions of gluinos \footnote{The corrections from the
Higgs interactions to the total width of $t \to b W$ are not unique
to the MSSM since they also exist in the general two-Higgs-doublet
models. Such corrections are found to be less than $1\%$ for
the whole parameter space \cite{Hollik}. Therefore we do not calculate
such corrections in this paper.}.
The SUSY-EW and SUSY-QCD corrections to the total width were first
investigated in \cite{Yang1,Yang2} without considering squark mixings,
and later were re-studied extensively in \cite{Garcia,Dabelstein,Brandenburg}
by taking into account the squark mixings. Although our aim in this paper is to study
the corrections to $\Gamma_L/\Gamma$, $\Gamma_-/\Gamma$ and
$\Gamma_+/\Gamma$, we will also repeat the calculations of the
corrections to the total width for two reasons:
 One of the reasons is
that the
previous SUSY-EW calculations were performed many years ago and some
SUSY parameter space viable at that time has been ruled out by
recent experiments.
The other reason is that the SUSY-EW corrections to the total width
were found to be exceptionally large ( exceeds $10\%$ in a large part
of parameter space)  \cite{Garcia} and should, therefore, be checked carefully.
In fact we find they are much smaller for reasons we will discuss.

This paper is organized as follows. In section II, we present
relevant formula for our calculations. In section III,  SUSY-EW
corrections are calculated and compared with previous studies.
In section IV, we examine SUSY-QCD corrections. For the numerical
calculations of both SUSY-EW and SUSY-QCD corrections, we first
perform a scan in the typical allowed parameter space to find out
the typical size of the corrections, and then
consider some special scenarios (such as a very light sbottom or
gluino) to figure out if the corrections are exceptionally
large in some regions of parameter space.
The conclusions are given in section V.

\section{\bf Formalism }

In this section we present the formulas for calculating new
physics contributions to $t \to b W$
    in the so-called "$G_F$-scheme" \cite{On-shell}
    where the Fermi constant $G_F$ and the
    pole masses $m_W$, $m_t$, $m_b$, $\cdots$ are chosen as input parameters.
    This on-shell scheme is the most convenient one for studying such new
    physics contributions.
These formulas are
valid for studying one-loop corrections to the decay $t \to  W b$
in any renormalizable new physics model.

The $t W b$ vertex at one-loop level takes the form\cite{On-shell}
\begin{eqnarray}
\Gamma_{\mu} = - i 2^{3/4} G_{F}^{\frac{1}{2}} m_W  V_{t b}
&&\left \{ \gamma_{\mu} P_L (1 + \frac{1}{2} \delta Z_b^L +
\frac{1}{2} \delta Z_t^L + \delta Z_1^W -\delta Z_2^W -\frac{1}{2}
\Pi_W
-\frac{1}{2} \Delta r^{\rm new} + F_L ) \right.  \nonumber \\
&& \left.+ \gamma_{\mu}  P_R  F_R  +  p_{t \mu } P_L  H_L + p_{t
\mu } P_R H_R  \right \} \ .\label{formula}
\end{eqnarray}
Here $P_{R,L}\equiv\frac{1}{2} \left ( 1 \pm \gamma_5 \right )$
are the chirality projectors. The form factors $F_{L,R}$ and
$H_{L,R}$ represent the contributions from the irreducible vertex
loops. (For the comparison of our results with Ref.~\cite{Garcia},
we adopted the same notations for the form factors as in
Ref.~\cite{Garcia}.) $\delta Z_b^L$ and  $\delta Z_t^L$ denote
respectively the field renormalization constant for bottom quark
and top quark. $\delta Z_{1,2}^W$ are the renormalization
constants for the $SU(2)$ gauge fields $W^a_{\mu}$ and the
coupling constant $g_2$, defined as $W^a_{\mu}\to (Z_2^W)^{1/2}
W^a_{\mu}$ and $g_2\to  Z_1^W (Z_2^W)^{-3/2} g_2$ with
$Z_{1,2}^W\equiv 1 + \delta Z_{1,2}^W$. Their difference is given
by \cite{On-shell}
\begin{eqnarray}
\delta Z_1^W-\delta Z_2^W=-\frac{1}{s_Wc_W} \frac{\Sigma_{\gamma Z}(0)}{m_Z^2} \ ,
\end{eqnarray}
where $c_W=\cos{\theta_W} $, $s_W=\sin{\theta_W}$, and
$\Sigma_{\gamma Z}(0)$ is the unrenormalized $\gamma$-Z transfer
at zero momentum. $\Pi_W$ in Eq.(\ref{formula}) represents the
finite wave function renormalization of the external $W$-boson and
is given by $\Pi_W=\Sigma_W^{\prime}(m_W^2)+\delta Z^W_2$, where
$\Sigma_W^\prime \equiv \partial \Sigma_W(p^2) /\partial p^2$ with
$\Sigma_W$ denoting the $W$-boson self-energy. $\Delta r^{\rm new}
$ is the corresponding new physics contribution to $\Delta r$,
which is the radiative correction to the relation of $M_W$, $M_Z$,
$\alpha$ (the fine structure constant) and $G_F$ defined by
\cite{On-shell}
\begin{eqnarray}
M_W^2\left (1-\frac{M_W^2}{M_Z^2}\right )
  =\frac{\pi \alpha} {\sqrt 2 G_F} \frac{1}{1-\Delta r} .
\end{eqnarray}

By decomposing $\Delta r^{\rm new} $ into the universal part and
non-universal part, i.e., $\Delta r^{\rm new} = \Delta r^{\rm
new}_{U} + \Delta r^{\rm new}_{NU} $, and inserting the explicit
forms of $\delta Z_{1,2}^W$, one can re-express  $\Gamma_{\mu}$
as \cite{On-shell}
\begin{eqnarray}
\Gamma_{\mu} &=& - i 2^{3/4} G_{F}^{\frac{1}{2}} m_W  V_{t b}
\left \{ \gamma_{\mu} P_L (1 + \frac{1}{2} \delta Z_b^L +
\frac{1}{2} \delta Z_t^L -\frac{1}{c_W s_W} \frac{\Sigma_{\gamma
Z}(0)}{m_Z^2} \right . \nonumber \\
&& \left . -\frac{1}{2} \Sigma_W^{\prime} (m_W^2) -\frac{1}{2}
\frac{\Sigma_W (0)- \delta m_W^2}{m_W^2} -\frac{1}{2} \Delta
r^{\rm new}_{NU} + F_L )+ \gamma_{\mu} P_R F_R + p_{t \mu } P_L  H_L +
p_{t \mu } P_R H_R \right \} \ .  \label{formula1}
\end{eqnarray}
In our calculations we adopted all the conventions in \cite{On-shell}
except for the fermion-field renormalization.
To be valid in general cases, the self-energy of a fermion is decomposed as
\begin{eqnarray}
\Sigma_f (p) =\Sigma_f^L (p^2) \p_slash  P_L+ \Sigma_f^R (p^2)
\p_slash P_R +m_f \Sigma_f^{S L} (p^2) P_L + m_f \Sigma_f^{S R}
(p^2) P_R \ .   \label{self1}
\end{eqnarray}
Using the on-shell conditions for external fermion fields \cite{Denner},
we obtain $\delta Z_f$ as
\begin{eqnarray}
\delta Z_f^L&= & \tilde{\Sigma }_f^L (m_f^2) +m_f^2 \left [
  \tilde{\Sigma }_f^{L \prime}(m_f^2)+ \tilde{\Sigma }_f^{R \prime }(m_f^2)
  + \tilde{\Sigma }_f^{S L \prime} (m_f^2)
  + \tilde{\Sigma }_f^{S R \prime} (m_f^2) \right ]\ ,   \label{self2} \\
\delta Z_f^R&=& \tilde{\Sigma }_f^R (m_f^2)
  +m_f^2 \left [\tilde{\Sigma }_f^{L \prime}(m_f^2)
  + \tilde{\Sigma}_f^{R \prime }(m_f^2)+ \tilde{\Sigma }_f^{S L \prime} (m_f^2)
  + \tilde{\Sigma}_f^{S R \prime} (m_f^2) \right ] \ , \label{self2a}
\end{eqnarray}
    where for $\tilde\Sigma$ one takes the real part of the loop integrals only
    but keeping all parts for the coupling constants in the fermion
    self-energies.

The rate of the top quark decay into a polarized $W$-boson can be
obtained by using the explicit expression of polarization vector
of $W$-boson\footnote{For the construction of covariant polarization vector of
$W$-boson, see Eq.(5.16) in \cite{Denner}.}  or using the
project technique introduced in \cite{Groot1,Groot2}.
Through some tedious calculations we obtain
\begin{eqnarray}
\Gamma_L =\frac{G_F m_W^2 m_t |V_{tb} |^2 }{8 \sqrt{2} \pi }
\frac{(1- x^2)^2}{x^2} &&\left \{  1+  Re ( \delta Z_b^L + \delta
Z_t^L + 2 \delta Z_1^W -2 \delta Z_2^W - \Pi_W - \Delta r^{\rm new} +
2 F_L ) \right .  \nonumber \\
&& \left . +  Re(H_R) m_t (1- x^2) \right \},  \label{gammal} \\
\Gamma_-=\frac{G_F m_W^2 m_t |V_{tb} |^2 }{8 \sqrt{2} \pi }
(1-x^2)^2   2 & & \left \{ 1+ Re ( \delta Z_b^L + \delta Z_t^L + 2 \delta
Z_1^W -2 \delta Z_2^W - \Pi_W - \Delta r^{\rm new} + 2 F_L ) \right \} ,
\label{gammam}
\end{eqnarray}
where $\Gamma_L $ ($\Gamma_- $) denotes the rate of the top quark decay into
longitudinal (transverse-minus ) $W$-boson and $x=m_W/m_t $. In
deriving  Eqs.(\ref{gammal},\ref{gammam}) we have neglected the $b$-quark
mass for simplicity and this will introduce an uncertainty of several
permille on $ \Gamma_{L,-}$. Another consequence of neglecting
$m_b$ is $\Gamma_+ =0$ due to angular momentum conservation
\cite{Groot1} and, as a result, the total decay rate of $t \to b W$
is obtained by $\Gamma=\Gamma_L+\Gamma_-$.
In our following calculations, we retain the bottom quark mass only when it appears in
couplings or in the sbottom mass matrix \cite{Haber} since in those cases
the bottom quark mass is multiplied by $\tan \beta$ and can not be
neglected for large $\tan \beta$.

We define the ratios
\begin{eqnarray}
\hat{\Gamma}_{L,-}= \Gamma_{L,-}/\Gamma,
\end{eqnarray}
which can be measured in experiments. In our numerical results we will
present the relative SUSY corrections to them, i.e.,
$\delta \hat{\Gamma}_{L,-}/\hat{\Gamma}_{L,-}^0$ with $\delta \hat{\Gamma}_{L,-}$
denoting the SUSY corrections and $\hat{\Gamma}_{L,-}^0$ denoting the SM predictions.

\section{SUSY electroweak corrections}
In this section we investigate the SUSY electroweak corrections.
First we consider vertex corrections and quark
self-energy corrections as depicted in Fig.~\ref{EWfeyman}. This part
of the corrections are expected to be sizable for large $\tan \beta$.
In Appendix A, we list in detail the relevant Feynman rules and
present  our analytical results. We checked that our results do reduce to those in
\cite{Yang1} when switching off squark mixings and neglecting the $b$-quark mass
in quark-squark-chargino (or neutralino) interactions.

The SUSY-EW contributions to the self-energies of gauge bosons,
which are necessary to calculate for $\delta Z_{1,2}^W $ and
$\Pi_W $, have been calculated by several groups
\cite{Yang1,Self-energy}. We recalculated them and find that our
results agree with those in \cite{Yang1} when switching off squark
mixings. In our numerical calculations we will take into account
the mixings between top-squarks and between sbottoms.

As to the SUSY-EW contributions to $\Delta r$, both the universal and
non-universal parts are nicely presented in \cite{Deltr}. We checked the
analytic expressions and incorporated them in our FORTRAN code.
Our numerical results show that the total contribution to the
width $\Gamma (t\to Wb)$ from $\delta Z_{1,2}^W$, $\Pi_W$ and
$\Delta r^{\rm SUSY-EW}$ is generally less than $1\%$.
\begin{figure}[hbt]
\begin{center}
\epsfig{file=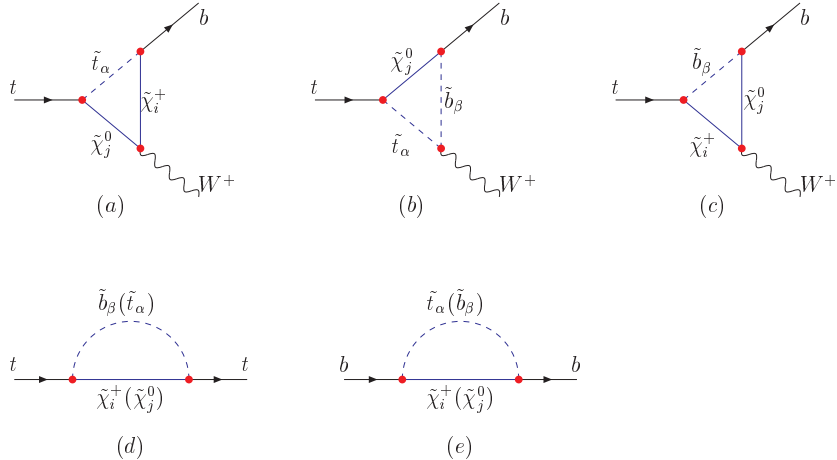,width=12cm}
\caption{ Some of the Feynman diagrams of SUSY-EW corrections to
$t \to Wb$: (a-c) are  vertex loops;
          (d,e) are top and bottom quark self-energy loops.} \label{EWfeyman}
\end{center}
\end{figure}
\vspace*{-1.0cm}
\begin{figure}[hbt]
\begin{center}
\epsfig{file=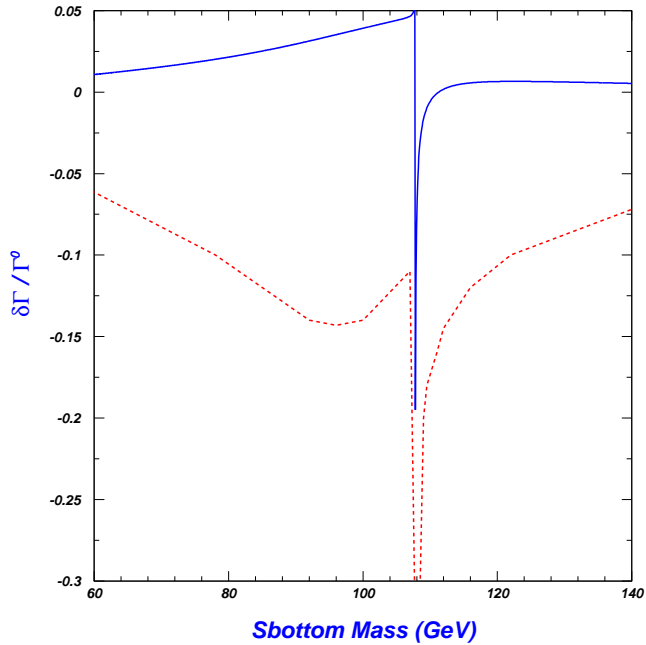,width=9cm} \vspace*{.5cm}
\caption{ SUSY electroweak corrections to the width $\Gamma(t \to W b)$ versus
          sbottom mass.  The solid curve is our result and the dashed curve is
          taken from Fig.~7 of Ref.[13]. The input parameters are those of
          case C in Fig.~7 in Ref.[13]. }
\label{Compare}
\end{center}
\end{figure}

As pointed out in Appendix A, our vertex corrections are quite different from those
in \cite{Garcia}. To see the numerical difference between
our results and those in \cite{Garcia}, we present a comparison in Fig.~\ref{Compare},
which corresponds to the case C of Fig.~7 in \cite{Garcia}.
We see from Fig.~\ref{Compare} that the difference is significant.
The results of \cite{Garcia} are much larger in magnitude than ours:
in a large part of parameter space the results of \cite{Garcia}
exceed $10\%$ in size while our results are below $5\%$.

Fig.~\ref{Compare} also shows  a common feature for both results,
namely, there exists a peak in the correction  at $m_{\tilde{b}}
\simeq 109$ GeV. This peak comes from the ``threshold effect"
\cite{Threshold}, which is explained in Appendix B.

Note that some of the input parameters in Fig.~\ref{Compare}, which were allowed
at the time when the calculations in \cite{Garcia} were performed,
may no longer be allowed nowadays. In our following numerical calculations,
we will consider the current constraints on the parameters.

The relevant SUSY parameters in our calculations are: (1) $\tan
\beta$, $\tilde{M}_{Q}$, $\tilde{M}_{U}$, $\tilde{M}_{D} $, $\mu$
and $A_{t,b}$ which enter the mass matrices of top-squark and sbottom
\cite{Haber} (see Appendix A); (2) soft-breaking gaugino masses
$\tilde{M}_1 $ and $\tilde{M}_2$ appearing in the chargino and
neutralino mass matrices (see Appendix A); (3) the masses of
sleptons and
  the squarks of the first two generations
involved in the
calculation of $\delta Z_{1,2}^W $, $\Pi_W$ and $\Delta r^{\rm
SUSY-EW}$. To lessen the number of the parameters in (3),
we assume an universal soft breaking parameter $\tilde{M}_L$
and neglect left-right mixings for sleptons and the squarks of
the first two generations. Such assumptions do not affect our numerical
results significantly since, as we said before, the contributions
from $\delta Z_{1,2}^W$, $\Pi_W$ and $\Delta r^{\rm new}$ are
small for the allowed parameter space. As for the soft breaking gaugino
mass parameters $\tilde{M}_1 $ and $\tilde{M}_2 $, we assume the
relation $\tilde{M}_1 \simeq 0.5 \tilde{M}_2$ \cite{GUT}. With
these assumptions the relevant SUSY parameters are reduced to
\begin{equation} \label{para}
\tilde{m}_{Q}, \tilde{m}_{U}, \tilde{m}_{D}, A_t, A_b, \mu, \tan{\beta}, \tilde{M}_L, \tilde{M}_2 .\label{input}
\end{equation}

To estimate the size of the SUSY-EW effects, we first
performed a scan over the nine parameters in Eq.~(\ref{input}).
In our scan we restricted the parameters with mass dimensions to be less
than $1$ TeV and considered the following experimental constraints:
\begin{itemize}
\item[{\rm(1)}] $\mu>0$ and a large $\tan\beta$ in the range $5\le\tan\beta\le 70$, which seems
                to be favored by the muon $g-2$ measurement~\cite{Brown01}.
\item[{\rm(2)}] $\delta \rho < 0.002 $\cite{PDG00}, which
                will constrain the mass splitting between stops and
                sbottoms. (We use the program FeynHiggs\cite{FeynHiggs}
                to generate the values of $\delta \rho $ and the Higgs masses.)
\item[{\rm(3)}] The LEP and CDF lower mass bounds on sparticles and the lightest
                CP-even Higgs boson \cite{PDG00,LEP}
\begin{eqnarray}
& & m_{{\tilde t}}\geq 86.4~GeV,~~ m_{{\tilde b}}\geq 75.0~GeV,
~~m_{\tilde{\chi}^+} \geq 67.7~GeV ,~~m_{\tilde{\chi}^0} \geq 40~GeV,
 \nonumber \\
 & & m_{\tilde{l}} \geq 95~GeV, ~~m_{\tilde{\nu}} \geq 41~ GeV,
~~m_{\tilde{q}} \geq 138~GeV, ~~m_{h^0} \ge 91~GeV ,\label{constrain}
\end{eqnarray}
where $m_{\tilde{q}}$ denotes the mass of squarks of the first two generations.
\end{itemize}
\vspace*{-0.5cm}
\begin{figure}[hbt]
\begin{center}
\epsfig{file=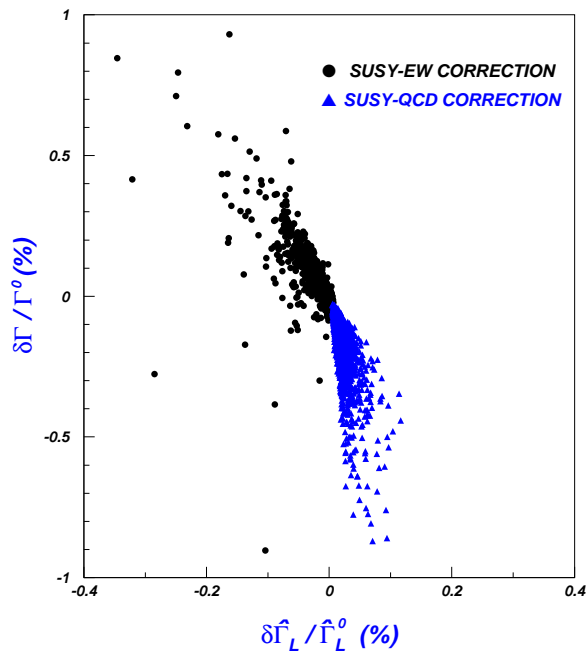,width=8cm}
\caption{The scattered plot in the plane of $\delta\Gamma/\Gamma^0$ versus
         $\delta\hat{\Gamma}_L/\hat{\Gamma}_L^0$. }  \label{Scan}
\end{center}
\end{figure}
Our scan results are shown in Fig.~\ref{Scan} in the plane of
$\delta \Gamma/\Gamma^0$ versus $\delta
\hat{\Gamma}_L/\hat{\Gamma}_L^0$. In our calculations throughout
this paper, we have fixed $G_F=1.16637 \times 10^{-5}$ GeV$^{-2}$,
$m_W=80.451$ GeV, $m_t =174.3$ GeV and $m_b =4.8$ GeV.
Fig.~\ref{Scan} shows that the SUSY-EW correction is generally
below one percent in size and tends to be positive for $\Gamma$
and negative for $\hat{\Gamma}_L$.

\begin{figure}[hbt]
\begin{center}
\epsfig{file=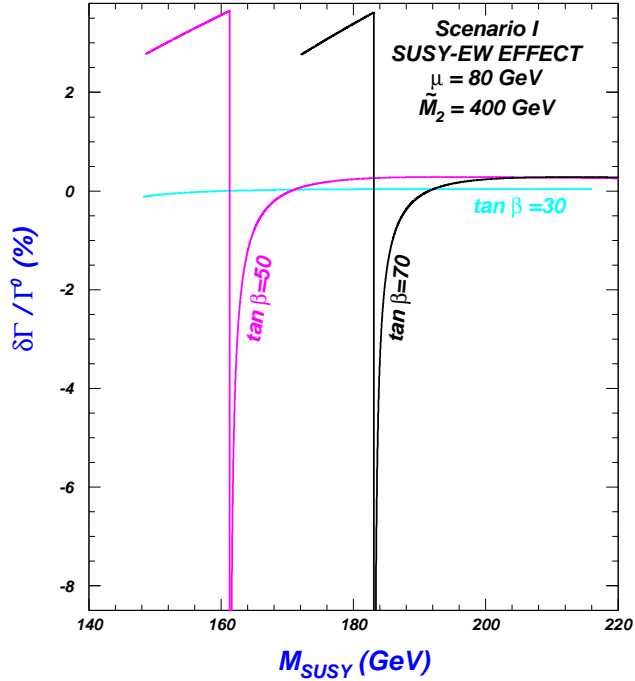,width=9cm}
\caption{SUSY electroweak corrections to $\Gamma$ versus $M_{SUSY}$ in Scenario I.}
\label{Ewdwidcase1}
\end{center}
\end{figure}

\begin{figure}[hbt]
\begin{center}
\epsfig{file=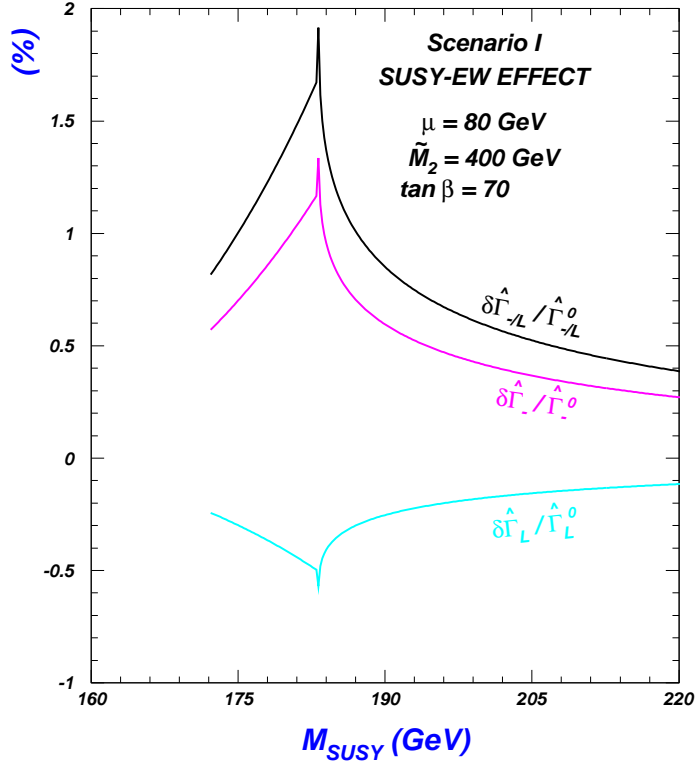,width=10cm}
\caption{SUSY eletroweak corrections to $\hat{\Gamma}_{-,L}$
         and $\hat{\Gamma}_{-/L}$($\equiv \hat{\Gamma}_-/\hat{\Gamma}_L$)
         versus $M_{SUSY} $ in Scenario I.}
\label{Ewdulcase1}
\end{center}
\end{figure}

Although the SUSY-EW correction turns out to be quite small (below
one percent in size) in the dominant part of the allowed parameter
space, we found from our scan that the correction can reach a few
percent in some narrow corners of the parameter space. These
corners have two feathers: (1) $\mu \ll \tilde{M}_2$ so that the
lighter chargino and the lightest neutralino are Higgsino-like,
and $\tan\beta$ is large so that the $b$-quark Yukawa couplings
in quark-squark-Higgsino interactions can be greatly
enhanced~\cite{Haber}; (2) The two sparticles in a self-energy
loop of top quark lie below the threshold, or in other words, the
sum of the masses of the two sparticles is lighter than top quark
mass. Generally speaking, for
$m_{\tilde{b_1}}+m_{\tilde{\chi}^+_1}<m_t$ the sbottom($\tilde
b_1$)-chargino($\tilde \chi^+_1$) loop can give a relatively large
contribution. With the increase of $m_{\tilde{b_1}}$ and
$m_{\tilde{\chi}^+_1}$, the contributions from this loop decrease
as required by a decoupling theorem. A peak in correction size
occurs at $m_{\tilde b_1}+m_{\tilde \chi^+_1}=m_t$, as shown in
Fig.~\ref{Compare} and explained in Appendix B. To show the size
of the corrections in such corners of the parameter space, we
present some numerical results in two special scenarios:
\begin{itemize}
\item {\rm Scenario I}:  We  assume $\mu \ll M_2$ and assume an universal soft-breaking mass $M_{SUSY}$
                         for squarks.
\item {\rm Scenario II}:  We assume a very light sbottom $\tilde{b}_1$ of about $5$ GeV.
    So far such a light sbottom has not been ruled out by current experiments
    if the sbottom mixing angle $\theta_b$ is tuned to satisfy
    $\cos\theta_b \simeq 0.38$ \cite{Carena}.
    For such a light sbottom, $\tan \beta$ must be large enough to
   cause a sufficiently large mass splitting between the two sbottom mass
   eigenstates. In addition, the lighter top-squark $\tilde{t}_1$ cannot
   be heavier than $300 $ GeV in order to satisfy the $\delta \rho$ bound.
\end{itemize}
The results in Scenario I are plotted in Figs.~\ref{Ewdwidcase1}
and \ref{Ewdulcase1}. We present our result only in the region
   where the constraints given in the paragraph following Eq.(\ref{input})
are satisfied. The peaks
in these figures happen at
$m_{\tilde{b}_1}+m_{\tilde{\chi}^+_1}=m_t$ where $m_{\tilde{b}_1}$
and $m_{\tilde{\chi}^+_1}$ are the lighter sbottom and top-squark
masses, respectively. Fig.~\ref{Ewdwidcase1} shows that below the
threshold the SUSY-EW corrections to the width can be as large as
$3\%$ for $\tan \beta \geq 50$ and after crossing the threshold,
the corrections change sharply to $-8\%$
and then quickly decrease
in size. Concerning the behavior at threshold, we want to point out
that the results within the region $0\leq m_{\tilde{b}} +
m_{\tilde{\chi}}-m_t < \Gamma/2$ are not reliable since
perturbative expansion of the $S$-matrix element breaks down for these
points \cite{Garcia} (also see discussions in Appendix B).
Fig.~\ref{Ewdulcase1} shows that $\delta
\hat{\Gamma}_-/\hat{\Gamma}_-^0$ can reach $1\%$ near the
threshold and its size is generally larger than that of
$\delta\hat{\Gamma}_L/\hat{\Gamma}_L^0$.
Now we turn to Scenario II. For numerical calculations in Scenario II, it is inconvenient to
take the parameters in Eq.(\ref{input}) as input. Instead, we may
choose
\begin{eqnarray} \label{parameter}
m_{\tilde{b}_1}, m_{\tilde{t}_1},m_{\tilde{t}_2}, \mu, A_b,
\theta_b, \tan \beta, \tilde{M}_L, \tilde{M}_2
\end{eqnarray}
as input parameters. Using the formula listed in Appendix A, one may
relate these two sets of parameters.
In our numerical calculations, we keep $\mu$ and $\tan\beta$ as variables
and study the dependence on them.
Other parameters are fixed to be\footnote{We found from numerical calculations
that the results are not sensitive to the parameters $m_{\tilde{t}_1}$,  $m_{\tilde{t}_2}$,
$\tilde{M}_L$ and $A_b$.}
\begin{eqnarray}
& & m_{\tilde{t}_1} =150~ {\rm GeV},
    ~~m_{\tilde{t}_2}=500~ {\rm GeV}, ~~\tilde{M}_L=A_b=300~ {\rm GeV}, \nonumber \\
& & m_{\tilde{b}_1}=5~ {\rm GeV},~~\cos\theta_b=0.38,~~\tilde{M}_2 =400~ {\rm GeV}.
\end{eqnarray}
Note that we fixed such values for $m_{\tilde{b}_1}$ and $\cos\theta_b$ since
this scenario is characterized by assuming
a very light sbottom $\tilde{b}_1$ of about $5$ GeV
with the sbottom mixing angle sytisfying $\cos\theta_b \simeq 0.38$
(see the definition of this scenario).

\begin{figure}[hbt]
\begin{center}
\epsfig{file=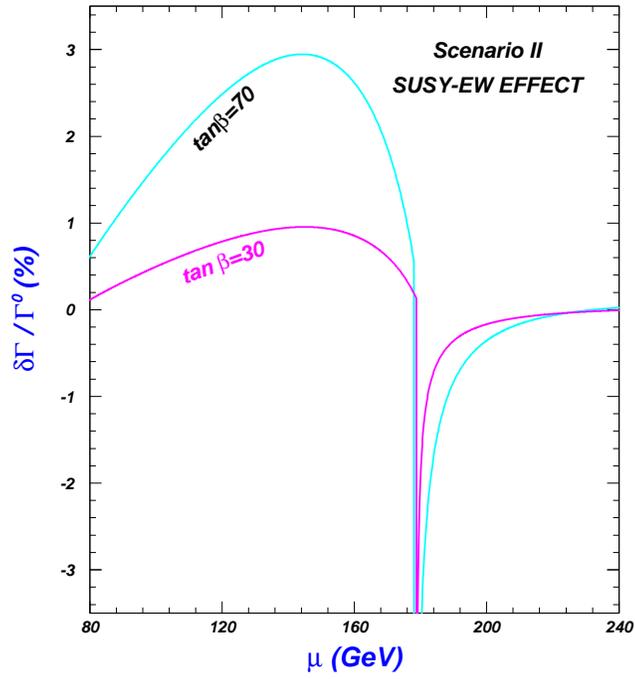,width=9cm}
\caption{SUSY electroweak corrections to $\Gamma$ versus $\mu$ in Scenario II.}
\label{Ewdwidcase2}
\end{center}
\end{figure}

\begin{figure}[hbt]
\begin{center}
\epsfig{file=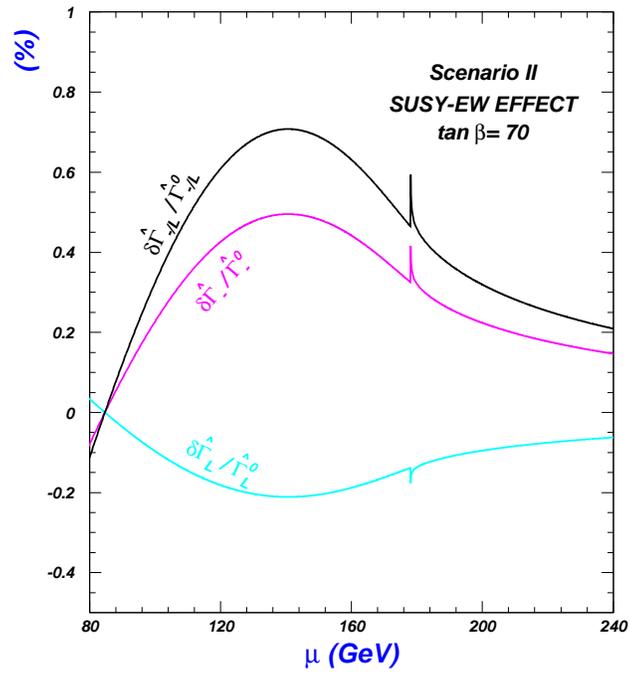,width=9cm}
\caption{SUSY electroweak corrections to $\hat{\Gamma}_{-,L}$ and  $\hat{\Gamma}_{-/L}$
         versus $\mu$ in Scenario II. }
\label{Ewdulcase2}
\end{center}
\end{figure}

The results in Scenario II are plotted in Figs.~\ref{Ewdwidcase2} and \ref{Ewdulcase2}.
In  Fig.~\ref{Ewdwidcase2} we plot $\delta \Gamma/ \Gamma^0$ versus $\mu$
and  in  Fig.~\ref{Ewdulcase2} we plot  $\delta\hat{\Gamma}_{-,L}/\hat{\Gamma}_{-,L}^0$
and  $\delta\hat{\Gamma}_{-/L}/\hat{\Gamma}_{-/L}^0$ versus $\mu$.
Again we see that a large $\tan \beta$ can cause large SUSY-EW
corrections.  Fig.~\ref{Ewdwidcase2} shows a broad region of $\mu$ in which
corrections to the width can be a few percent. Like in Fig.~\ref{Ewdwidcase1} the peaks
occur at $m_{\tilde{b}_1}+m_{\tilde{\chi}}^+ =m_t$.
Fig.\ref{Ewdulcase2} shows that, like in Scenario I, the SUSY-EW corrections to
$\hat{\Gamma}_{-}$ and $\hat{\Gamma}_L $ are of opposite sign and the
size of the former is larger than that of the latter.

\section{\bf SUSY-QCD corrections}
\begin{figure}[hbt]
\begin{center}
\epsfig{file=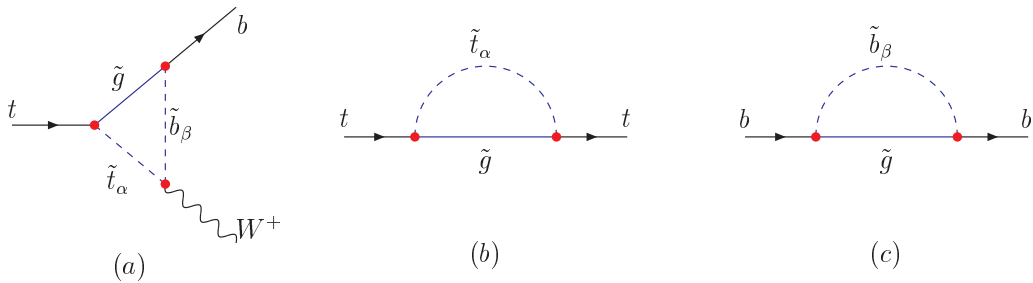,width=16cm}
\caption{Feynman diagrams for SUSY-QCD contribution to $t \to  W b$.}
\label{fey-qcd}
\end{center}
\end{figure}

In this section we investigate SUSY-QCD effects. Compared with
SUSY-EW corrections, SUSY-QCD corrections are easier to calculate
since $\delta Z_{1,2}^W $, $\Pi_W $ and  $\delta r^{\rm new} $ receive
no contribution from SUSY-QCD interaction at the one-loop level and
consequently, one only needs to calculate quark self-energy and
vertex correction depicted in Fig.~\ref{fey-qcd}.
Our analytical results for the form factors $F_{L,R}$ and $H_{L,R}$ agree with
those in Ref.\cite{Dabelstein} and thus we do not present
our analytic results here.
Our numerical results for the corrections to the total width
$\Gamma(t \to  W b)$  agree with Ref.\cite{Brandenburg} for the
same choice of parameters.

The parameters involved in numerical evaluations are the gluino mass
$m_{\tilde g}$, and the parameters $\tilde{m}_{Q}$, $\tilde{m}_{U}$,
$\tilde{m}_{D}$, $A_t$, $A_b$, $\mu$ and $\tan{\beta}$ which enter
the mass matrices of top-squark and sbottom\cite{Haber}. In order
to compare the size of SUSY-QCD corrections with that of SUSY-EW
corrections, we have shown our scan results of SUSY-QCD
corrections already in Fig.~\ref{Scan} in the preceding section. In our
scan we have considered the relevant constraints
given in the paragraph following Eq.(\ref{input})
and we also required $m_{\tilde{g}} \geq 200$ GeV,
which is the current experimental bound for the gluino in the
framework of the minimal supergravity model\cite{PDG00}. From
Fig.~\ref{Scan} one sees that the SUSY-QCD corrections and SUSY-EW
corrections tend to have opposite signs and thus may
cancel in large parts of parameter space. This differs drastically
from the conclusion of Ref.\cite{Garcia,Dabelstein} which claims
that the two type corrections have the same sign and thus the
combined contribution can reduce the width $\Gamma (t \to W b)$ by
as much as $25\%$.

    We see from Fig.~\ref{Scan} that in the allowed parameter space with
    $m_{\tilde g} > 200$ GeV the SUSY-QCD corrections are smaller than $1\%$
    in magnitude.
However, we found that the corrections can reach a few percent in some
corners of the parameter space if we relax the
 bound  $m_{\tilde{g}} \geq 200$ GeV (this bound is model-dependent).
In the following we present some numerical results by relaxing this bound
$m_{\tilde{g}} \geq 200$ GeV, and in order
to compare with SUSY-EW corrections we consider the same two scenarios
as in the preceding section.

\begin{figure}[hbt]
\begin{center}
\epsfig{file=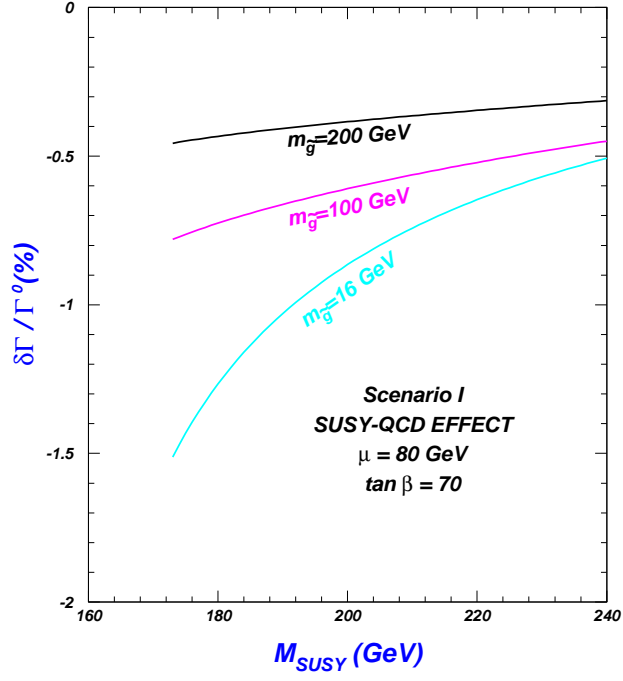,width=9cm}
\caption{SUSY-QCD correction to $\Gamma$ versus $M_{SUSY} $ in Scenario I.}
\label{qcddwidcase1}
\end{center}
\end{figure}

\begin{figure}[hbt]
\begin{center}
\epsfig{file=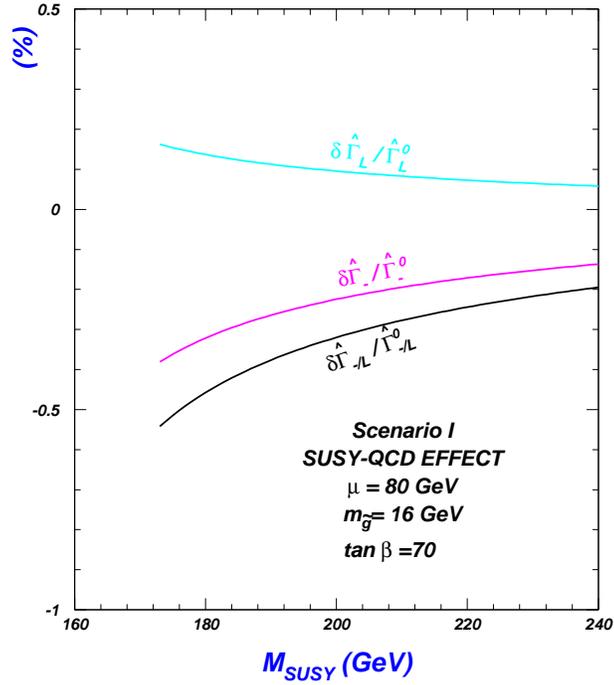,width=9cm}
\caption{SUSY-QCD correction to $\hat{\Gamma}_{-,L}$ and  $\hat{\Gamma}_{-/L}$ versus $M_{SUSY} $ in Scenario I.}
\label{qcddulcase1}
\end{center}
\end{figure}

\begin{figure}[hbt]
\begin{center}
\epsfig{file=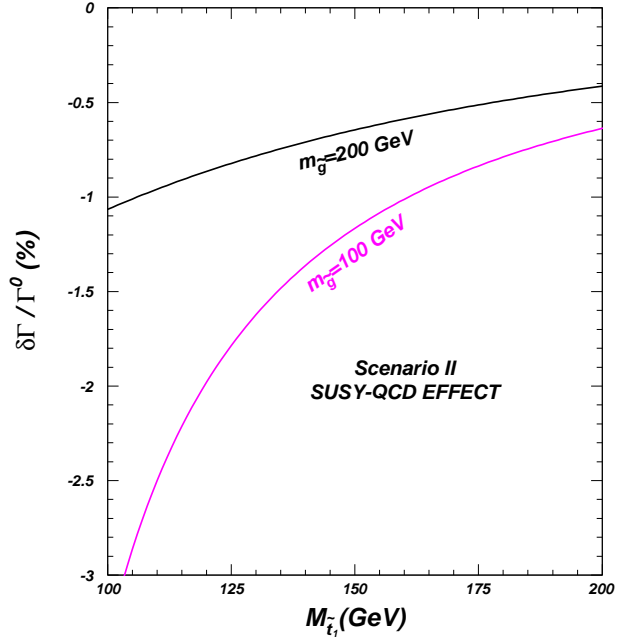,width=8.5cm}
\caption{SUSY-QCD corrections to $\Gamma$ verus  the lighter top-squark mass in Scenario II.}
\label{qcddwidcase2}
\end{center}
\end{figure}

\begin{figure}[hbt]
\begin{center}
\epsfig{file=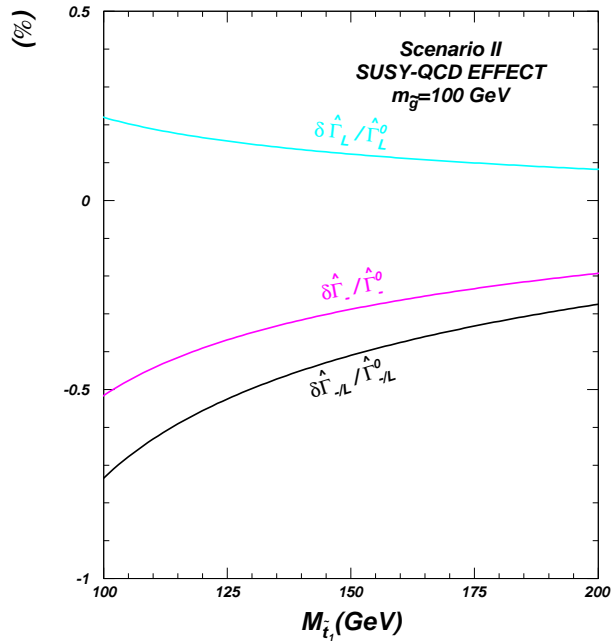,width=8.5cm}
\caption{SUSY-QCD corrections to $\hat{\Gamma}_{-,L}$ and  $\hat{\Gamma}_{-/L}$
         versus the lighter top-squark mass in Scenario II.}
\label{qcddulcase2}
\end{center}
\end{figure}

The results in Scenario I are plotted in Figs.~\ref{qcddwidcase1}
and \ref{qcddulcase1} for different $m_{\tilde{g}}$ values in the
allowed region of $M_{SUSY}$. We fixed $\mu=80$ GeV and $\tan\beta
=70$, and included the possibility of a very light gluino of 16
GeV, which has not been excluded by current experiments
\footnote{Such a very light gluino has not been excluded by
experiments and was recently used to solve the long-standing
$b$-quark production puzzle \cite{Berger}. To achieve that, a light
sbottom of about 5 GeV is also required, which leads to a stronger
constraint from $R_b$ on the SUSY parameters~\cite{Cao}.}
 (for comparison we also plot a curve with a 100 GeV
gluino although it may be severely disfavored by existing
experiments). Fig.~\ref{qcddwidcase1} shows that the magnitude of
SUSY-QCD corrections decrease as $M_{SUSY}$ or $m_{\tilde{g}}$
becomes large. Comparing with the SUSY-EW corrections in
Fig.~\ref{Ewdwidcase1}, one sees that the maximum size of the SUSY-QCD
corrections is smaller than that of the SUSY-EW corrections, which can
be greatly enhanced by the $b$-quark Yukawa coupling for large
$\tan \beta$.
Next we present results in Scenario II.
Similar to the SUSY-EW case, in Scenario II it is convenient to choose
\begin{eqnarray}
m_{\tilde{b}_1}, m_{\tilde{t}_1},m_{\tilde{t}_2}, \mu, A_b,
\theta_b, \tan \beta, m_{\tilde{g}}
\end{eqnarray}
as input parameters. Our numerical studies show that
the SUSY-QCD corrections are only sensitive to $m_{\tilde{t}_1}$
and $ m_{\tilde{g}}$. So we fix $m_{\tilde{b}_1} =5$ GeV,
$m_{\tilde{t}_2}=500$ GeV, $\mu =80$ GeV, $A_b=300$ GeV, $\tan \beta =70$
and $\cos \theta_b =0.38$ and plot the corrections in
Figs.~\ref{qcddwidcase2} and \ref{qcddulcase2} for different gluino masses.
Since in this scenario we assume a very light sbottom ($m_{\tilde{b}_1} =5$ GeV),
we do not consider the possibility of a very light gluino ($m_{\tilde{g}} =16$ GeV)
due to the $R_b$ constraints~\cite{Cao}.
From the figures we see that for a
$m_{\tilde{t}_1} =100$ GeV and $m_{\tilde{g}} =100$ GeV, SUSY-QCD
corrections can reach $-3\% $, comparable in magnitude with
SUSY-EW corrections.
\vspace*{-0.6cm}

\section{Summary and conclusion}
We investigated the one-loop SUSY-QCD and SUSY-EW corrections to
the top quark decay into a $b$-quark and a longitudinal or
transverse $W$-boson. The corrections are presented in terms of
the longitudinal ratio $\Gamma(t\to W_L b)/\Gamma(t\to W b)$ and
the transverse ratio $\Gamma(t\to W_- b)/\Gamma(t\to W b)$.
In order to compare with the existing results in the literature,
we also present the corrections to the total width $\Gamma(t\to W
b)$. To find out the typical size of these corrections, we
performed a scan in the typical allowed parameter space. In
order to decide if the corrections are exceptionally large in
certain corners of the parameter space, we then considered some special
scenarios (such as a very light sbottom or gluino). Our
observations are:
\begin{itemize}
\item[(1)] In most of the parameter space, both SUSY-QCD and SUSY-EW corrections to the
           ratios are less than $1\%$ in magnitude and they tend to have opposite signs.
           Only in some corners of the parameter space the corrections can reach a few percent.
\item[(2)] The corrections to the total width $\Gamma(t\to W b)$ agree well with previous calculations for
           the SUSY-QCD part, but differ significantly from previous calculations for the SUSY-EW part.
           Unlike the previous studies, which showed a large SUSY-EW correction
           of more than $10\%$ in magnitude for a large part of the parameter space,
           our SUSY-EW results are only a few percent at most.
\end{itemize}
We conclude by remarking that
these SUSY corrections, despite their smallness in size, should be taken into
account in the future precision tests of the top quark decay
$t\to W b$ to determine whether SUSY is indeed the new physics chosen by Nature.
(As far as we know, the
corrections to  the $tWb$ coupling in other kinds of new physics models
are also very small \cite{twb-other}.)
However, from the viewpoint of probing SUSY through revealing its
effects in the coupling $tWb$, the decay $t\to W b$ may not be the
best channel.  Some single top quark production channels  proceeding
through the  $tW b$ coupling may be complementary or even do better in
this aspect. For example, SUSY effects in the $tWb$ coupling may
cause observable effects in single top quark production in hadron
collisions~\cite{susy-singletop}, and the single top quark production
via electron-photon collision at a future linear collider is also
quite sensitive to anomalous $tWb$ coupling~\cite{nlc-singletop}.

\section*{Acknowledgment}
This work is supported in part by the Chinese Natural Science Foundation
and by the US Department of Energy, Division
of High Energy Physics under grant No. DE-FG02-91-ER4086.

\appendix
\section{SUSY-EW loop results }
\label{app-a}

In this appendix we first list the mass matrices of sfermions, charginos and
neutralinos involved in our calculations, then write down the relevant interaction
Lagrangian and finally present the explicit expressions of the
SUSY-EW contributions from the loop diagrams shown in Fig.\ref{EWfeyman}.

\subsection {Mass matrices of sfermions, charginos and neutralinos}

Assuming no generation mixing for squarks in the soft-breaking terms,
the mass-square matrices for up-type and down-type squarks in
each generation are given respectively by \cite{Haber}
\begin{eqnarray}
M_{\tilde u}^2& =& \left ( \begin{array}{cc} \tilde{M}_{Q}^2 + m_Z^2
       \cos{2 \beta} (\frac{1}{2}-e_u
       \sin^2{\theta_W})+m_u^2 & m_u ( A_u -\mu \cot \beta )   \\
        m_u ( A_u -\mu \cot \beta ) & \tilde{M}_{U}^2 + m_Z^2
        \cos{2 \beta}  e_u  \sin^2{\theta_W}+m_u^2
           \end{array}  \right ) \ , \label{su} \\
M_{\tilde d}^2& = &\left ( \begin{array}{cc} \tilde{M}_Q^2 - m_Z^2
         \cos{2 \beta} (\frac{1}{2}+ e_d
         \sin^2{\theta_W})+m_d^2 & m_d ( A_d -\mu \tan \beta )   \\
         m_d ( A_d -\mu \tan \beta ) & \tilde{M}_D^2 + m_Z^2
          \cos{2 \beta}  e_d  \sin^2{\theta_W}+m_d^2
           \end{array}  \right ) \ ,  \label{sd}
\end{eqnarray}
where $\tilde M_Q$, $ \tilde M_{U}$ and $\tilde M_D$ are
soft-breaking mass terms for left-handed squark doublet,
right-handed up and down squarks, respectively. $A_u$ ($A_d$) is
the coefficient of  the trilinear term $H_2 \tilde Q \tilde U$
($H_1 \tilde Q \tilde D $) in soft-breaking terms, and
$\tan\beta=v_2/v_1$ is the ratio of the vacuum expectation values of
the two Higgs doublets.

For the first two generation squarks, the left-right mixings can be
neglected  due to the smallness of relevant quark masses and thus
the weak-eigenstates are mass-eigenstates.
From Eqs.(\ref{su},\ref{sd}), we then get the relation between
the left-handed up-type squark mass $M_{\tilde u_L}$ and
the left-handed down-type squark mass $M_{\tilde d_L}$
\begin{eqnarray}
M_{\tilde u_L}^2=M_{\tilde d_L}^2+m_Z^2 \cos{2 \beta} (1-\sin^2\theta_W )+m_u^2-m_d^2 \ .
\end{eqnarray}

For the third generation squarks, the left-right mixings cannot
be neglected and the mass-eigenstates $\tilde{q}_{1,2}$ ($q=t,b$)
are related to the weak-eigenstates $\tilde{q}_{L,R}$ by an unitary
rotation $R$
\begin{eqnarray}
\left ( \begin{array}{l} \tilde{q}_1 \\  \tilde{q}_2 \end{array} \right )
 = R \left ( \begin{array}{l} \tilde{q}_L \\  \tilde{q}_R \end{array} \right )
   \equiv \left ( \begin{array}{cc} \cos \theta_{q} & \sin \theta_{q}  \\
           -\sin \theta_{q} & \cos \theta_{q} \end{array}  \right )
   \left ( \begin{array}{l} \tilde{q}_L \\  \tilde{q}_R \end{array} \right ) \ ,
\end{eqnarray}
where $\theta_{q}$ is the so-called mixing angle between  $\tilde{q}_{L}$
and  $\tilde{q}_{R}$.

The mass-square matrices in Eqs.(\ref{su}, \ref{sd}) can be alternatively
expressed by the squark masses and mixing angle as
\begin{eqnarray}
M^2_{\tilde q} =\left ( \begin{array}{cc} \cos^2 \theta_{q}
m_{\tilde{q}_1}^2 +\sin^2 \theta_{q} m_{\tilde{q}_2}^2   & \sin
\theta_{q} \cos \theta_q (m_{\tilde{q}_1}^2- m_{\tilde{q}_2}^2)
\\ \sin \theta_{q} \cos \theta_q (m_{\tilde{q}_1}^2- m_{\tilde{q}_2}^2)  &
\sin^2 \theta_{q} m_{\tilde{q}_1}^2 +\cos^2 \theta_{q}
m_{\tilde{q}_2}^2
           \end{array}  \right ) \ . \label{mixing}
\end{eqnarray}
By comparing Eq.(\ref{mixing}) with Eqs.(\ref{su}, \ref{sd}), one
can obtain the relationship between the two set of input
parameters in the squark sector, ($m_{\tilde q_1}$, $m_{\tilde q_2}$,
$\theta_q$) and ($\tilde{M}_{Q} $, $\tilde{M}_{U} $,
$\tilde{M}_D$, $A_{t,b}$, $\mu $). The mass matrices for sleptons
take the similar forms as for squarks. But due to the smallness of
lepton masses, the left-right mixings can be neglected for the
three generations.

The mass matrices of charginos and neutralinos are also involved in
our calculations.
The chargino mass matrix can be diagonalized by two unitary matrices
$U $ and $V$
\begin{eqnarray}
U^{\ast} \left ( \begin{array}{cc} \tilde{M}_2 & m_W \sqrt{2} \sin
\beta \\ m_W \sqrt{2} \cos{\beta} & \mu \end{array} \right )
V^{-1} ={\rm diag} \left \{  m_{\tilde{\chi}^+_i} \right \}
\ \ \ \ (i=1,2) \ .
\label{charg}
\end{eqnarray}
The neutralino mass matrix is given by
\begin{eqnarray}
Y = \left ( \begin{array}{cccc} \tilde{M}_1 & 0 & -m_Z \sin
\theta_W \cos \beta & m_Z \sin \theta_W \sin \beta \\ 0 &
\tilde{M}_2 & m_Z \cos \theta_W \cos \beta & - m_Z \cos \theta_W
\sin \beta \\ -m_Z \sin \theta_W \cos \beta &m_Z \cos \theta_W
\cos \beta & 0 & -\mu
\\ m_Z \sin \theta_W \sin \beta & - m_Z \cos \theta_W \sin \beta
& -\mu & 0 \end{array} \right ) \ ,\label{neutra}
\end{eqnarray}
and it  can be  diagonalized by an unitary matrix $N$
\begin{eqnarray}
N^{\ast} Y N^{-1} ={\rm diag} \left \{ m_{\tilde{\chi}^0_j} \right
\}\ \ \ \ (j=1,2,3,4) \ . \label{Ndef}
\end{eqnarray}
In the above expressions $\tilde{M}_1$ and $\tilde{M}_2$ are
respectively the soft-breaking $U(1)$ and $SU(2)$ gaugino masses.
$m_{\tilde{\chi}^+_{i}} $ and $m_{\tilde{\chi}^0_j}$ are the
masses of charginos and neutralinos, respectively. One point we
want to mention is that $U $, $ V $ and $N$ defined in Eq.(\ref{charg},
\ref{Ndef}) are not unique and one should choose $U$, $ V$ and
$N$ in a way that all the masses of charginos and neutralinos are positive.

\subsection{Interaction Lagrangian}

The interaction Lagrangian relevant for our calculations is given by
\begin{eqnarray}
{\cal L} & =&  -\frac{g}{2} \left \{ \tilde{t}^{\ast}_{\alpha}
     \bar{\tilde{\chi}}^-_i (A_{\alpha i}^{t} -B_{\alpha i}^t \gamma_5) b
   +\tilde{b}^{\ast}_{\beta} \bar{\tilde{\chi}}^+_i (A_{\beta i}^{b}
   -B_{\beta i}^b \gamma_5 ) t  + \tilde{t}^{\ast}_{\alpha}
   \bar{\tilde{\chi}}^0_j (A_{\alpha j}^{t (0)} -B_{\alpha j}^{t (0)}\gamma_5 ) t
    +\tilde{b}^{\ast}_{\beta} \bar{\tilde{\chi}}^0_j (A_{\beta j}^{b (0)}
    -B_{\beta j}^{b (0)} \gamma_5 ) b \right \} \nonumber \\
& & -\frac{g}{\sqrt{2}} R_{\alpha 1}^t R_{\beta 1}^b \tilde{t}^{\ast}_{\alpha}
    \stackrel{\leftrightarrow}{\partial^{\mu}} \tilde{b}_{\beta}
    W_{\mu}^+ + g \bar{\tilde{\chi}}^0_j \gamma^{\mu} (O_{ij}^L P_L
    + O_{ij}^R P_R ) \tilde{\chi}^-_i W_{\mu}^+  \nonumber \\
& & +\frac{g^2}{2} (R^t_{\alpha_1 1} R^t_{\alpha_2 1}
    \tilde{t}^{\ast}_{\alpha_1} \tilde{t}_{\alpha_2}
    + R^b_{\beta_1 1} R^b_{\beta_2 1} \tilde{b}^{\ast}_{\beta_1} \tilde{b}_{\beta_2} )
      W_{\mu}^- W^{\mu + } + h.c. \ ,   \label{Lagrag}
\end{eqnarray}
where the sum over repeated indices is implied, and
\begin{eqnarray}
A^t_{\alpha i} &= & R^t_{\alpha 1} (V^{\ast}_{i 1}
  -\lambda_b U_{i2})-R^t_{\alpha 2} \lambda_t V_{i2}^{\ast} \ ,\label{At} \\
B^t_{\alpha i} &= & R^t_{\alpha 1} (V^{\ast}_{i 1} +\lambda_b U_{i2})
  -R^t_{\alpha 2} \lambda_t V_{i2}^{\ast} \ ,\label{Bt} \\
A^b_{\beta i} &= & R^b_{\beta 1} (U^{\ast}_{i 1} -\lambda_t V_{i2})
  -R^b_{\beta 2} \lambda_b U_{i2}^{\ast} \ ,\\
B^b_{\beta i} &= & R^b_{\beta 1} (U^{\ast}_{i 1} +\lambda_t V_{i2})
  -R^b_{\beta 2} \lambda_b U_{i2}^{\ast} \ ,\\
A^{t(0)}_{\alpha j} & =& R^t_{\alpha 1} (\lambda_t N_{j4}
  +\sqrt{2} (\frac{1}{6} N_{j1}^{\ast} \tan \theta_W
  +\frac{1}{2}N_{j2}^{\ast} )) + R^t_{\alpha 2} ( \lambda_t N_{j4}^{\ast}
  -\sqrt{2} \frac{2}{3} N_{j1} \tan \theta_W ) \ ,   \label{At0} \\
B^{t(0)}_{\alpha j} & =& R^t_{\alpha 1} (-\lambda_t N_{j4}
  +\sqrt{2} (\frac{1}{6} N_{j1}^{\ast} \tan \theta_W
  +\frac{1}{2}N_{j2}^{\ast} )) + R^t_{\alpha 2} ( \lambda_t N_{j4}^{\ast}
  +\sqrt{2} \frac{2}{3} N_{j1} \tan \theta_W ) \ , \label{Bt0} \\
A^{b(0)}_{\beta j} & =& R^b_{\beta 1} (\lambda_b N_{j3}
  +\sqrt{2}(\frac{1}{6} N_{j1}^{\ast} \tan \theta_W -\frac{1}{2}N_{j2}^{\ast} ))
  +R^b_{\beta 2} ( \lambda_b N_{j3}^{\ast}
  +\frac{\sqrt{2}}{3} N_{j1} \tan \theta_W ) \ ,\\
B^{b(0)}_{\beta j} & =& R^b_{\beta 1} (-\lambda_b N_{j3}
  +\sqrt{2}(\frac{1}{6} N_{j1}^{\ast} \tan \theta_W
  -\frac{1}{2}N_{j2}^{\ast} )) + R^b_{\beta 2} ( \lambda_b N_{j3}^{\ast}
  -\frac{\sqrt{2}}{3} N_{j1} \tan \theta_W ) \ , \\
O_{ij}^L & =& -\frac{1}{\sqrt{2}} N_{j4} V_{i2}^{\ast} + N_{j2}V_{i1}^{\ast} \ ,\\
O_{ij}^R & =& \frac{1}{\sqrt{2}} N_{j3}^{\ast} U_{i2} + N_{j2}^{\ast} U_{i1} \ .
\label{ORij}
\end{eqnarray}
Here $\lambda_t=\frac{m_t}{\sqrt{2} m_W \sin \beta}$ and
$\lambda_b =\frac{m_b}{\sqrt{2} m_W \cos \beta}$.
    As for the notations for chargino and neutralino mass matrices, we follow
    the notations of Ref.\cite{Haber} rather than those of Ref.\cite{Garcia}.
    Note, however,
    that the resulting interactions are the same.

\subsection{Analytic results of SUSY-EW loops}

To compare our results with those in Ref.~\cite{Garcia}, we
present our results in the same way as in Ref.\cite{Garcia}.
First, from matrices in Eqs.(\ref{At}-\ref{ORij}) we define
the following matrices
\begin{eqnarray}
& & A_{\pm}^t = A^t\pm B^t\ , \ \ \ \ A_{\pm}^{t (0)}  = A^{t (0)} \pm B^{t (0)} \ , \\
& & A_{\pm}^b = A^b \pm B^b\ , \ \ \ \ A_{\pm}^{b (0)} = A^{b (0)} \pm B^{b (0)} \ ,
\end{eqnarray}
and construct the combinations as
\begin{eqnarray}
A^{(1)}&=&-A_+^{t \ast} O^R A_-^{t (0)}\ , \ \ \ \ E^{(1)}=-A_-^{t \ast}O^R A_-^{t (0)}\ , \\
B^{(1)}&=&-A_+^{t \ast} O^R A_+^{t (0)}\ , \ \ \ \ F^{(1)}=-A_-^{t \ast} O^R A_+^{t (0)}\ , \\
C^{(1)}&=&-A_+^{t \ast} O^L A_-^{t (0)}\ , \ \ \ \ G^{(1)}=-A_-^{t \ast} O^L A_-^{t (0)}\ , \\
D^{(1)}&=&-A_+^{t \ast} O^L A_+^{t (0)}\ , \ \ \ \ H^{(1)}=-A_-^{t \ast} O^L A_+^{t (0)} \ ,\\
A^{(2)}&=&-R_{\alpha 1}^t R_{\beta 1}^b A_+^{b (0) \ast} A_-^{t(0)}\ , \ \ \ \
        C^{(2)}=-R_{\alpha 1}^t R_{\beta 1}^b A_-^{b(0) \ast} A_-^{t (0)} \ ,\\
B^{(2)}&=&-R_{\alpha 1}^t R_{\beta 1}^b A_+^{b (0) \ast} A_+^{t(0)} \ ,
      \ \ \ \ D^{(2)} =  -R_{\alpha 1}^t R_{\beta 1}^b A_-^{b(0) \ast} A_+^{t (0)} \ , \\
A^{(3)}&=&A_+^{b (0) \ast} O^L A_-^b \ ,\ \ \ \ E^{(3)} =A_-^{b(0) \ast} O^L A_-^b \ ,\\
B^{(3)}&=&A_+^{b (0) \ast} O^L A_+^b \ ,\ \ \ \ F^{(3)} =A_-^{b(0) \ast} O^L A_+^b \ ,\\
C^{(3)}&=&A_+^{b (0) \ast} O^R A_-^b \ ,\ \ \ \ G^{(3)} =A_-^{b(0) \ast} O^R A_-^b \ ,\\
D^{(3)}&=&A_+^{b (0) \ast} O^R A_+^b \ ,\ \ \ \ H^{(3)} =A_-^{b(0) \ast} O^R A_+^b \ .
\end{eqnarray}
Then the contributions of vertex loop diagrams (a,b,c) in Fig.\ref{EWfeyman}
to the form factors $F_{L,R}$ and $H_{L,R}$ are given as follows.
\begin{itemize}
\item {\bf Diagram (a)}:
\begin{eqnarray}
F_L^{(a)} &=& \frac{\sqrt{2} g^2}{4 V_{tb}} \frac{1}{16 \pi^2} \left
    \{ -\frac{1}{2} D^{(1)} (-1 +4 C_{24}) +m_{\tilde{\chi}^0_j}
    m_{\tilde{\chi}^-_i} B^{(1)} C_0 -m_{\tilde{\chi}^0_j} m_t C^{(1)}
    (C_{11}-C_{12})  \right . \nonumber \\
& & \left .+m_{\tilde{\chi}^-_i} m_t A^{(1)} (C_0+ C_{11}-C_{12})
   -m_t^2 D^{(1)} (C_{11}-C_{12}+C_{21}-C_{23}) -m_W^2 D^{(1)}
   (C_{22}-C_{23}) \right \} \ , \\
F_R^{(a)} & = & \frac{\sqrt{2} g^2}{4 V_{tb}} \frac{1}{16 \pi^2}
   \left\{ -\frac{1}{2} E^{(1)} (-1 +4 C_{24}) +m_{\tilde{\chi}^0_j}
   m_{\tilde{\chi}^-_i} G^{(1)} C_0 -m_{\tilde{\chi}^0_j} m_t F^{(1)}
   (C_{11}-C_{12})  \right . \nonumber \\
& & \left .+m_{\tilde{\chi}^-_i} m_t H^{(1)} (C_0+ C_{11}-C_{12})
   -m_t^2 E^{(1)} (C_{11}-C_{12}+C_{21}-C_{23}) -m_W^2 E^{(1)}
   (C_{22}-C_{23}) \right \} \ ,\\
H_L^{(a)} & = & \frac{\sqrt{2} g^2}{4 V_{tb}} \frac{1}{16 \pi^2}
  \left \{ 2 m_{\tilde{\chi}^0_j} F^{(1)} (C_{11}-C_{12})
  +2m_{\tilde{\chi}^-_i}H^{(1)} C_{12}
  +2m_t E^{(1)} (C_{11}-C_{12}+C_{21}-C_{23})  \right \} \ , \\
H_R^{(a)} & = & \frac{\sqrt{2} g^2}{4 V_{tb}} \frac{1}{16 \pi^2}
   \left \{ 2 m_{\tilde{\chi}^0_j} C^{(1)} (C_{11}-C_{12}) +2
   m_{\tilde{\chi}^-_i}A^{(1)} C_{12}
   +2m_t D^{(1)} (C_{11}-C_{12}+C_{21}-C_{23})  \right \} \ ,
\end{eqnarray}
where the functions $C_0$ and $C_{nm}$ are 3-point Feynman integrals
defined in Ref.~\cite{Axelrod} with functional dependence
$C_0, C_{nm} (-p_t, p_W,m_{\tilde{t}_\alpha},m _{\tilde{\chi}_j^0},
m_{\tilde{\chi}_i^-})$.

\item {\bf Diagram (b)}:
\begin{eqnarray}
F_L^{(b)} &= & \frac{g^2}{2 V_{tb}} \frac{1}{16 \pi^2} \left \{ -B^{(2)} C_{24}\right \} \ ,\\
F_R^{(b)} &= & \frac{g^2}{2 V_{tb}} \frac{1}{16 \pi^2} \left \{ -C^{(2)} C_{24} \right \} \ ,\\
H_L^{(b)} &= & \frac{g^2}{2 V_{tb}} \frac{1}{16 \pi^2}
        \left \{m_{\tilde{\chi}^0_j} D^{(2)} (C_0+C_{11}) +m_t C^{(2)}
                (C_{12}-C_{11}-C_{21}+C_{23}) \right \} \ ,\\
H_R^{(a)} &= & \frac{g^2}{2 V_{tb}} \frac{1}{16 \pi^2} \left \{
         m_{\tilde{\chi}^0_j} A^{(2)} (C_0+C_{11}) +m_t B^{(2)}
         (C_{12}-C_{11}-C_{21}+C_{23}) \right \} \ ,
\end{eqnarray}
with the Feynman integrals $C_0, C_{nm} (-p_t, p_W, m_{\tilde{\chi}^0_j},
                 m_{\tilde{t}_{\alpha}}, m_{\tilde{b}_{\beta}} )$.

\item {\bf  Diagram (c) }:

    $F_{L,R}^{(c)}$ and $H_{L,R}^{(c)}$
                     can be obtained from the corresponding
                     $F_{L,R}^{(a)}$ and $H_{L,R}^{(a)}$
                     with the replacements:
\begin{eqnarray}
m_{\tilde{t}_{\alpha}} \to m_{\tilde{b}_{\beta}}\ , \ \ \ \ \ \
     m_{\tilde{\chi}_j^0} \leftrightarrow m_{\tilde{\chi}_i^-}\ , \ \ \ \ \ \
     A^{(1)} \to A^{(3)}\ ,  \ \ \ \ \ \
     B^{(1)} \to B^{(3)}\ , \nonumber \\
C^{(1)} \to C^{(3)}\ , \ \ \ \ \ \ D^{(1)} \to D^{(3)} \ ,\ \ \ \ \ \
E^{(1)} \to E^{(3)}\ , \ \ \ \ \ \ H^{(1)} \to H^{(3)} \ .
\end{eqnarray}
\end{itemize}
For the self-energy loop diagrams in (d,e) of Fig.~\ref{EWfeyman},
we only present the corresponding forms of
$\Sigma_{\tilde{\chi}^-}$ (from chargino loops) and of
$\Sigma_{\tilde{\chi}^0}$ (from neutralino loops). The
renormalization constants $\delta Z_{t,b}^L$ can be obtained
in a straitforward manner by using Eqs.(\ref{self2}, \ref{self2a}). The
results are  given by
\begin{itemize}
\item  {\bf  Diagram (d) }:
\begin{eqnarray}
\Sigma_{\tilde{\chi}^-}^t (p)&=&  \frac{g^2}{4} \frac{1}{16 \pi^2}
   \left \{  ( |A^b_+|^2 \p_slash P_L +|A^b_-|^2  \p_slash P_R )
   B_1(p, m_{\tilde{\chi}_i^-}, m_{\tilde{b}_{\beta}}) \right . \nonumber \\
& & \left . - m_{\tilde{\chi}_i^-} ( A_-^{b \ast} A_+^b P_L
     +A_+^{b\ast} A_-^b P_R ) B_0 (p, m_{\tilde{\chi}_i^-}, m_{\tilde{b}_{\beta}} )
     \right \} \ , \label{topself1} \\
\Sigma_{\tilde{\chi}^0}^t (p)&=&  \frac{g^2}{4} \frac{1}{16 \pi^2}
     \left \{ (|A^{t (0)}_+|^2 \p_slash P_L +|A^{t (0)}_-|^2 \p_slash P_R )
              B_1 (p, m_{\tilde{\chi}_j^0}, m_{\tilde{t}_{\alpha}}) \right . \nonumber \\
& & \left . -m_{\tilde{\chi}_j^0} ( A_-^{t (0) \ast} A_+^{t (0)} P_L
    +A_+^{t(0) \ast} A_-^{t (0)} P_R ) B_0 (p, m_{\tilde{\chi}_j^0}, m_{\tilde{t}_{\alpha}})
   \right \} \ . \label{topself2}
\end{eqnarray}

\item {\bf  Diagram (e) }:
\begin{eqnarray}
\Sigma_{\tilde{\chi}^-}^b (p)&= & \frac{g^2}{4} \frac{1}{16 \pi^2}
  \left \{ ( |A^t_+|^2  \p_slash P_L +|A^t_-|^2  \p_slash P_R )
  B_1(p, m_{\tilde{\chi}_i^-}, m_{\tilde{t}_{\alpha}}) \right . \nonumber \\
& & \left . - m_{\tilde{\chi}_i^-} ( A_-^{t \ast} A_+^t P_L
   + A_+^{t\ast} A_-^t P_R ) B_0 (p, m_{\tilde{\chi}_i^-},m_{\tilde{t}_{\alpha}} )\right \}\ ,  \\
\Sigma_{\tilde{\chi}^0}^b (p)&= & \frac{g^2}{4} \frac{1}{16 \pi^2}
  \left \{ (  |A^{b (0)}_+|^2  \p_slash P_L
  +|A^{b (0)}_-|^2  \p_slash P_R ) B_1 (p, m_{\tilde{\chi}_j^0},
   m_{\tilde{b}_{\beta}}) \right . \nonumber \\
& & \left . -m_{\tilde{\chi}_j^0} ( A_-^{b (0) \ast} A_+^{b (0)} P_L
    + A_+^{b(0) \ast} A_-^{b (0)} P_R ) B_0 (p, m_{\tilde{\chi}_j^0},
      m_{\tilde{b}_{\beta}} ) \right \} \ .
\end{eqnarray}
\end{itemize}
Comparing our results with those in \cite{Garcia}, we found
that our self-energy loop results essentially agree with those of
\cite{Garcia}, but our vertex  loop results disagree with  those
of \cite{Garcia}. The analytic results of \cite{Garcia} seem to
have some explicit errors, e.g.,
it seems to be impossible to cancel the UV divergence in $F_L$,
terms like $m_t C_0 $ should not appear in
$H_{L,R} $ and $G_2$ and $G_3$ should be interchanged.

\section{ Explanation of threshold behaviors}
\label{app-b} The two-point loop functions $B_{0,1}^{\prime}(p,
m_1, m_2) |_{p^2=m_t^2}$ enter our results via $\delta Z_t^L$. In
order to explain the threshold behavior in Figs.(\ref{Compare},
\ref{Ewdwidcase1}, \ref{Ewdwidcase2}), we study the behavior of
$B_{0,1}^{\prime}(m_0, m_1, m_2)$ near the threshold point
$m_0=m_1+m_2$.  We take $B_0^{\prime}$ as an example, the
behavior of $B_1^{\prime}$ is similar.

For fixed $m_0$ and $m_2$, $B_0^{\prime} (m_0, m_1, m_2)$ as a
function of $m_1$ can be expressed as
\begin{eqnarray}
B_0^{\prime} &= & \int_0^1 \frac{x -x^2}{m_0^2 x^2 -(m_0^2 +m_1^2
   -m_2^2) x+m_1^2} d x  \nonumber \\
&=&\left \{ \begin{array}{lll}
    \frac{1}{m_0^2} \int_0^1 \frac{x-x^2}{(x-a)^2-b^2} dx \ \ \ \
              & {\rm for }& \ \ m_1 < m_0-m_2 \ ,\\
    \frac{1}{m_0^2} \int_0^1 \frac{x-x^2}{(x-a)^2} dx \ \ \ \
              & {\rm for }& \ \ m_1=m_0-m_2 \ , \\
   \frac{1}{m_0^2} \int_0^1 \frac{x-x^2}{(x-a)^2 + b^2} dx \ \ \ \
              & {\rm for }& \ \ m_1 > m_0-m_2 \ ,
\end{array} \right . \label{B0p}
\end{eqnarray}
where
\begin{eqnarray}
a&=& \frac{m_0^2+m_1^2-m_2^2}{2 m_0^2} \ , \\
b&=& \frac{1}{2
m_0^2}\sqrt{|(m_0+m_1+m_2)(m_0+m_1-m_2)(m_0-m_1+m_2)(m_0-m_1-m_2)|}\
.
\end{eqnarray}
From Eq.(\ref{B0p}), one may infer that, in regions below
threshold ($m_1<m_0-m_2$) or above threshold ($m_1>m_0-m_2$),
$B_0^{\prime}$ is an continuous function of $m_1$. At the
threshold point ($m_1=m_0-m_2$), $B_0^{\prime}$ is divergent due
to the fact $0<a< 1$. (In some calculation programs, such as
LoopTools \cite{hahn} used in our calculations, an imaginary part
is added to the propagators in the loops to avoid the divergence
at the threshold point.)

In the limit $m_1\to m_0-m_2$ from the lower (i.e. $m_1<m_0-m_2$),
one has $B_0^{\prime}\to -2+(1-2a)\ln \left [(1-a)/a \right ]$ which is a
finite value and corresponds to the top points in
Figs.(\ref{Compare}, \ref{Ewdwidcase1}, \ref{Ewdwidcase2}).
 But in the limit $m_1\to m_0-m_2$ from the upper (i.e. $m_1>m_0-m_2$),
one has $B_0^{\prime}\to \infty$ which corresponds to the bottom point in
Figs.(\ref{Compare}, \ref{Ewdwidcase1}, \ref{Ewdwidcase2}). Since these
bottom points tend to take very large negative values, the perturbative calculation
is not reliable at these points and a special treatment is needed \cite{threshold}.

\begingroup\raggedright
\endgroup


\begin{thebibliography}{99}

\bibitem{sensitive} For example, top-pair production at hadron collisions are
                    quite sensitive to new physics.
               For comprehensive model-independent analyses,
               see, e.g.,
               C. T. Hill and S. J. Parke, \PRD49, 4454 (1994);
               K. Whisnant, {\it et al.},  \PRD56, 467 (1997);
               K. Hikasa, {\it et al.}, \PRD58, 114003 (1998).

\bibitem{review} For recent reviews on top quark, see, e.g.,
               C. T. Hill and E. Simmons, hep-ph/0203079;
               C.-P. Yuan,  hep-ph/0203088;
               E. Simmons, hep-ph/0211335;
               S. Willenbrock, hep-ph/0211067;
               D. Chakraborty, J. Konigsberg, D. Rainwater, hep-ph/0303092.

\bibitem{new-production} See, e.g., A. Datta, {\it et al.}, \PRD56, 3107 (1997);
                  R. J. Oakes,  {\it et al.}, \PRD57, 534 (1998);
                  K. Hikasa, J. M. Yang, B.-L. Young, \PRD60, 114041 (1999);
                  C. Balazs, H.-J. He, C.-P. Yuan, \PRD60, 114001  (1999);
                  H.-J. He, C.P. Yuan, \PRL83, 28 (1999);
                  G. Burdman,  \PRL83, 2888 (1999);
                  P. Chiappetta,  {\it et al.}, \PRD61, 115008 (2000);
                  J. Cao, Z. Xiong and J. M. Yang, \NPB651, 87 (2003); \PRD67, 071701 (2003).

\bibitem{new-decay} See, e.g., M. Hosch,  {\it et al.},  \PRD58, 034002 (1998);
                    S. Mrenna and C.P. Yuan, \PLB367, 188 (1996).
                    C.~S.~Li, R.~J.~Oakes and J.~M.~Yang, \PRD49, 293 (1994);
                    J.~L.~Lopez, D.~V.~Nanopoulos and R.~Rangarajan, \PRD56, 3100  (1997);
                    G. Eilam,  {\it et al.}, \PLB510, 227 (2001);
                    K.J. Abraham,  {\it et al.}, \PLB514, 72 (2001); \PRD63, 034011 (2001);
                    T. Han and J. Hewett, \PRD60, 074015 (1999);
                    F. del Aguila, J. A. Aguilar-Saavedra, R. Miquel, \PRL82, 1628 (1999);
                    X.~L. Wang  {\it et al.}, \PRD50, 5781 (1994).

\bibitem{Nelson} See, e.g., C. A. Nelson, {\it et. al.}, \PRD56, 5928(1997);
                 \PRD57, 5923 (1998);  C. A. Nelson and A. M. Cohen,
                 \EPJC8, 393 (1999); C. A. Nelson and L. J. Adler,
                 \EPJC17, 399 (2000); C. A. Nelson, \EPJC19, 323 (2001);
                  F. del Aguila and J. A. Aguilar-Saavedra, \PRD67, 014009 (2003).

\bibitem{Bigi} Y.~Bigi, {\it et. al.}, \PLB181, 157 (1986).

\bibitem{CDF}  CDF collaboration, T.~Affolder, {\it et al., } \PRL84, 216 (2000).

\bibitem{Willenbrock} S. Willenbrock , Rev.~Mod.~Phys.~{\bf 72}, 1141(2000).

\bibitem{Groot1} M.~Fischer, S.~Groote, J.~G.~K\"orner and M.~C.~Mauser, \PLB451, 406 (1999);
                 \PRD63, 031501 (2001); \PRD65, 054036 (2002).

\bibitem{Groot2} M.~Fischer, S.~Groote, J.~G.~K\"orner and M.~C.~Mauser, \PRD67, 091501 (2003).

\bibitem{Yang1} J.~M.~Yang and C.~S.~Li, \PLB320, 117 (1994).

\bibitem{Yang2} C.~S.~Li, J.~M.~Yang and B.~Q.~Hu, \PRD48, 5425 (1993).

\bibitem{Garcia} D.~Garcia, R.~A.~Jim\'enez, J.~Sol\'a  and W.~Hollik, \NPB427, 53 (1994).

\bibitem{Dabelstein} A.~Dabelstein, W.~Hollik, C.~J\"unger, R.~A.~ Jim\'enez and J.~Sol\'a,
                     \NPB454, 75 (1995).

\bibitem{Brandenburg} A.~Brandenburg and M.~Maniatis, \PLB545, 139 (2002).

\bibitem{Hollik} B.~Grzadkowski and W.~Hollik, \NPB384, 101 (1992).

\bibitem{On-shell} M.~Bohm, H.~Spiesberger and W.~Hollik, Fortsch.~Phys.~{\bf 34}, 687 (1986);
                   W.~Hollik, Fortsch.~Phys.~{\bf 38}, 165(1990).
\bibitem{Denner} A.~Denner, Fortsch.~Phys.~{\bf 41}, 307 (1993).

\bibitem{Haber} H.~E.~Haber and G.~L.~Kane, Phys.~Rept.~{\bf 117}, 75 (1985);
                J.~F.~Gunion and H.~E.~Haber, \NPB272, 1 (1986).

\bibitem{Self-energy} See, e.g.,  J.~A.~Grifols and J.~Sol\'a, \NPB253, 47 (1985);
                      D.~Garcia and J.~Sol\'a, Mod.\ Phys.\ Lett.\ A {\bf 9}, 211 (1994);
                      R.~Hempfling and B.~A.~Kniehl, Z.\ Phys.\ C {\bf 59}, 263 (1993).

\bibitem{Deltr} P.~H.~Chankowski, {\it et.al.}, \NPB417, 101 (1994).

\bibitem{Threshold} J.~Fleischer and F.~Jegerlehner, \PRD23, 2001 (1981); \NPB216, 469 (1983);
                    A.~Dabelstein and W.~Hollik, Z.\ Phys.\ C {\bf 53}, 507 (1992).

\bibitem{GUT} H.~P.~Nilles, Phys.\ Rept.\ {\bf 110}, 1 (1984);
              A.~B.~Lahanas and D.~V.~Nanopoulos, Phys.\ Rept.\  {\bf 145}, 1 (1987).

\bibitem{Brown01} H.~N~Brown,~{\em et~al.}, Mu g-2 Collaboration, \PRL86, 2227 (2001).

\bibitem{PDG00} Particle Physics Group. \EPJC15, 274 (2000).

\bibitem{FeynHiggs} S.~Heinemeyer, W.~Hollik and G.~Weiglein,
                    Comput.\ Phys.\ Commun.\ {\bf 124}, 76 (2000).

\bibitem{LEP}  R.~Barate {\it et al.}, ALEPH Collaboration, \PLB499, 53 (2001);
               LEP Higgs Working Group, hep-ex/0107029; hep-ex/0107030;
              U.~Schwickerath, hep-ph/0205126.

\bibitem{Carena} M.~Carena, S.~Heinemeyer, C.~E.~Wagner and G.~Weiglein, \PRL86, 4463 (2001).

\bibitem{Berger} E.~L.~Berger, {\it et. al.}, \PRL86, 4231 (2001).

\bibitem{Cao} J.~Cao, Z.~Xiong and J.~M.~Yang, \PRL88, 111802 (2002);
              G.~C.~Cho, \PRL89, 091801 (2002);
              K.~Cheung and W.~Y.~Keung, \PRL89, 221801 (2002);
              Z. Luo and  J. L. Rosner, hep-ph/0306022;
              R. Malhotra, hep-ph/0306183.

\bibitem{twb-other} See, e.g., N.~Mahajan, hep-ph/0304235.

\bibitem{susy-singletop} C.~S.~Li,  {\it et al.}, \PRD57,2009 (1998); \PLB98, 298 (1997); \PRD55, 5780 (1997);

\bibitem{nlc-singletop}  J. J. Cao et al., \PRD58, 094004 (1998);
                         E. Boos, et al., \EPJC21, 81 (2001).

\bibitem{Axelrod} A.~Axelrod, \NPB209, 349 (1982).

\bibitem{hahn}  T.~Hahn, Nucl.~Phys.~Proc.~Suppl.{\bf 89}, 231 (2000);
                    Acta Phys.~Plon.~B{\bf 30}, 3469 (1999).

\bibitem{threshold} T. Bhattacharya and  S. Willenbrock, \PRD47, 4022 (1993);
                     B. A. Kniehl,  C. P. Palisoc and A. Sirlin, \NPB591, 296 (2000).
\end{thebibliography}
\end{document}